\documentclass[preprint,fleqn]{cas-dc}

 \usepackage[numbers]{natbib}

\usepackage{mathrsfs}

\usepackage{algorithm}
\usepackage{algpseudocode}
\usepackage{amsmath}
\makeatletter
\algrenewcommand\ALG@beginalgorithmic{\footnotesize}
\makeatother

\usepackage{subcaption}

\usepackage{longtable}

\usepackage{graphicx}
\usepackage{stfloats}
\usepackage{multicol}
\usepackage{array}
\usepackage{ragged2e}
\usepackage[edges]{forest}
\usetikzlibrary{shadows,arrows.meta, shadows.blur} 
\tikzset{
  parent/.style={align=center,text width=4.5cm,fill=gray!50,rounded corners=2pt},
  child/.style={align=center,text width=3cm,fill=gray!20,rounded corners=6pt},
  grandchild/.style={fill=white,text width=3cm}
}
\usepackage{colortbl,color}
\definecolor{gray}{rgb}{0.90, 0.90, 0.90}
\definecolor{ORANGE}{rgb}{0.94, 0.86, 0.51}
\definecolor{BLUE}{rgb}{0.54, 0.81, 0.94}
\definecolor{GREEN}{rgb}{0.67, 0.88, 0.69}
\newcolumntype{g}{>{\columncolor{Gray}}}
\newcolumntype{O}{>{\columncolor{ORANGE}}}
\newcolumntype{B}{>{\columncolor{BLUE}}}
\newcolumntype{V}{>{\columncolor{GREEN}}}
\usepackage{mathtools}

\usepackage{pifont}

\usepackage{braket}

\usepackage{soul}

\usepackage{xcolor}

\def\tsc#1{\csdef{#1}{\textsc{\lowercase{#1}}\xspace}}
\tsc{WGM}
\tsc{QE}
\tsc{EP}
\tsc{PMS}
\tsc{BEC}
\tsc{DE}

\usepackage{acro}
\DeclareAcronym{RIC}{short=RIC, long=RAN Intelligent Controller}
\DeclareAcronym{RAN}{short=RAN, long=Radio Access Network}
\DeclareAcronym{O-RAN}{short=O-RAN, long=Open RAN}
\DeclareAcronym{DU}{short=DU, long=Distributed Unit}
\DeclareAcronym{CU}{short=CU, long=Central Unit}
\DeclareAcronym{CU-UP}{short=CU-UP, long=CU- User Plane}
\DeclareAcronym{CU-CP}{short=CU-CP, long=CU- Control Plane}
\DeclareAcronym{RU}{short=RU, long=Radio Unit}
\DeclareAcronym{AI/ML}{short=AI/ML, long=Artificial Intelligence/Machine Learning}
\DeclareAcronym{eMBB}{short=eMBB, long=Enhanced Mobile Broadband}
\DeclareAcronym{URLLC}{short=URLLC, long=Ultra Reliable Low Latency Communications}
\DeclareAcronym{mMTC}{short=mMTC, long=massive Machine Type Communications}
\DeclareAcronym{eCPRI}{short=eCPRI, long=enhanced Common Public Radio Interface}
\DeclareAcronym{TLS}{short=TLS, long=Transport Layer Security}
\DeclareAcronym{IPsec}{short=IPsec, long=IP security}
\DeclareAcronym{MACsec}{short=MACsec, long=MAC security}
\DeclareAcronym{SMO}{short=SMO, long=Service Management and Orchestration}
\DeclareAcronym{MLB}{short=MLB, long=Mobility Load Balancing}
\DeclareAcronym{UE}{short=UE, long=User Equipment}
\DeclareAcronym{MRO}{short=MRO, long=Mobility Robustness Optimization}
\DeclareAcronym{KPI}{short=KPI, long=Key Performance Indicator}
\DeclareAcronym{IDS}{short=IDS, long=Intrusion Detection System}
\DeclareAcronym{SLA}{short=SLA, long=Service Level Agreement}
\DeclareAcronym{MLops}{short=MLops, long=ML operations}
\DeclareAcronym{E2SM}{short=E2SM, long=E2 Service Model}
\DeclareAcronym{E2AP}{short=E2AP, long=E2 Application Protocol}
\DeclareAcronym{SON}{short=SON, long=Self-Organized Networks}
\DeclareAcronym{SDN}{short=SDN, long=Software-Defined Networking}
\DeclareAcronym{3GPP}{short=3GPP, long=3rd Generation Partnership Project}
\DeclareAcronym{O-RAN-SC}{short=O-RAN-SC, long=O-RAN Software Community}
\DeclareAcronym{WG}{short=WG, long=Working Group}
\DeclareAcronym{RRM}{short=RRM, long=Radio Resource Management}
\DeclareAcronym{RLF}{short=RLF, long=Radio Link Failure}
\DeclareAcronym{VM}{short=VM, long=Virtual Machine}
\DeclareAcronym{NFV}{short=NFV, long=Network Function Virtualization}
\DeclareAcronym{V2X}{short=V2X, long=Vehicle to Everything}
\DeclareAcronym{NDT}{short=NDT, long=Network Digital Twin}
\DeclareAcronym{MITM}{short=MITM, long=Man-in-the-middle}
\DeclareAcronym{UTRAN}{short=UTRAN, long=Universal Terrestrial RAN}
\DeclareAcronym{E-UTRAN}{short=E-UTRAN, long=Evolved UTRAN}
\DeclareAcronym{RL}{short=RL, long=Reinforcement Learning}
\DeclareAcronym{MDP}{short=MDP, long=Markov Decision Problem}
\DeclareAcronym{NS}{short=NS, long=Network Slicing}
\DeclareAcronym{FL}{short=FL, long=Federated Learning}
\DeclareAcronym{Near-RT-RIC}{short=Near-RT RIC, long=Near-Real-Time RIC}
\DeclareAcronym{Non-RT-RIC}{short=non-RT RIC, long=non-Real-Time RIC}
\DeclareAcronym{Near-RT}{short=Near-RT, long=Near-Real-Time}
\DeclareAcronym{Non-RT}{short=non-RT, long=non-Real-Time}
\DeclareAcronym{IoT}{short=IoT, long=Internet of Things}
\DeclareAcronym{LSTM}{short=LSTM, long= Long short-term memory}
\DeclareAcronym{RAT}{short=RAT, long=Radio Access Technology}
\DeclareAcronym{NR}{short=NR, long=New Radio}
\DeclareAcronym{VNF}{short=VNF, long=Virtual Network Function}
\DeclareAcronym{PRB}{short=PRB, long=Physical Resource Block}
\DeclareAcronym{DoS}{short=DoS, long=Denial-of-Service}
\DeclareAcronym{DPI}{short=DPI, long=Deep Packet Inspection}
\DeclareAcronym{UAV}{short=UAV, long=Unmanned Aerial Vehicle}
\DeclareAcronym{MNO}{short=MNO, long=Mobile Network Operator}
\DeclareAcronym{5GC}{short=5GC, long=5G Core}
\DeclareAcronym{FCAPS}{short=FCAPS, long= {Fault/Config/Accounting/Performance/Security}}
\DeclareAcronym{FH}{short=FH, long=Fronthaul (Connects the DU with the RU)}
\DeclareAcronym{NSA}{short=NSA, long=Non-Standalone}
\DeclareAcronym{SA}{short=SA, long=Standalone}
\DeclareAcronym{NG-RAN}{short=NG-RAN, long=Next-Generation RAN}
\DeclareAcronym{NIST}{short=NIST, long=National Institute of Standards and Technology}
\DeclareAcronym{PCA}{short=PCA, long=Principal Component Analysis}
\DeclareAcronym{FGSM}{short=FGSM, long=Fast Gradient Sign Method} 
\DeclareAcronym{PGD}{short=PGD, long=Projected Gradient Descent}
\DeclareAcronym{C-W}{short=C\&W, long=Carlini and Wagner}
\DeclareAcronym{JSMA}{short=JSMA, long=Jacobian-based Saliency Map Attack}
\DeclareAcronym{DNN}{short=DNN, long=Deep Neural Network}
\DeclareAcronym{PINN}{short=PINN, long=Physics-Informed Neural Network}
\DeclareAcronym{LIME}{short=LIME, long=Local Interpretable Model-agnostic Explanations}
\DeclareAcronym{SHAP}{short=SHAP, long=Shapley Additive Explanations}
\DeclareAcronym{A1}{short=A1, long= Connects the non-RT RIC with the Near-RT RIC}
\DeclareAcronym{E2}{short=E2, long= Connects the Near-RT RIC with the E2 nodes}
\DeclareAcronym{X2/Xn}{short=X2/Xn, long= Connects the gNB with other eNBs/gNBs}
\DeclareAcronym{O1}{short=O1, long= Connects the SMO with the O-RAN for FCAPS}
\DeclareAcronym{E1}{short=E1, long= Connects the CU-CP with the CU-UP}
\DeclareAcronym{F1}{short=F1, long= Connects the CU with the DU (Midhaul)}
\DeclareAcronym{O-eNB}{short=O-eNB, long= O-RAN enabled eNB}
\DeclareAcronym{O2}{short=O2, long= Connects the O-cloud with the SMO}
\DeclareAcronym{Y1}{short=Y1, long= Service interface for consumers}
\DeclareAcronym{NG}{short=NG, long= Connects the NG-RAN with the 5GC}
\DeclareAcronym{xHaul}{short=xHaul, long= Transport network backhaul/midhaul/fronthaul}
\DeclareAcronym{ANN}{short=ANN, long= Artificial Neural Network}
\DeclareAcronym{CNN}{short=CNN, long= Convolutional Neural Network}
\DeclareAcronym{MAC}{short=MAC, long= Medium Access Control}

\begin{document}
\sloppy
\let\WriteBookmarks\relax
\def\floatpagepagefraction{1}
\def\textpagefraction{.001}
\shorttitle{Misconfiguration in O-RAN: Analysis of the impact of AI/ML}
\shortauthors{N.M. Yungaicela-Naula et~al.}

\title [mode = title]{Misconfiguration in O-RAN: Analysis of the impact of AI/ML}                      



\author[1]{Noe M. Yungaicela-Naula}[type=editor,
                        auid=000,bioid=1,
                        orcid=0000-0002-3131-0672]
\cormark[1]
\ead{n.yungaicela@qub.ac.uk}


\address[1]{Centre for Secure Information Technologies (CSIT),
Queen's University Belfast,
Belfast, BT3 9DT, Northern Ireland, UK}

\author[1]{Vishal Sharma,}[style=chinese]

\author[1]{Sandra Scott-Hayward}[%
   ]





\cortext[cor1]{Corresponding author}


\begin{abstract}
User demand on network communication infrastructure has never been greater with applications such as extended reality, holographic telepresence, and wireless brain-computer interfaces challenging current networking capabilities. \ac{O-RAN} is critical to supporting new and anticipated uses of 6G and beyond. It promotes openness and standardisation, increased flexibility through the disaggregation of \ac{RAN} components, supports programmability, flexibility, and scalability with technologies such as \ac{SDN}, \ac{NFV}, and cloud, and brings automation through the \ac{RIC}. Furthermore, the use of xApps, rApps, and \ac{AI/ML} within the {RIC} enables efficient management of complex {RAN} operations. However, due to the open nature of {O-RAN} and its support for heterogeneous systems, the possibility of misconfiguration problems becomes critical. In this paper, we present a thorough analysis of the potential misconfiguration issues in {O-RAN} with respect to integration and operation, the use of {SDN} and {NFV}, and, specifically, the use of {AI/ML}. The opportunity for {AI/ML} to be used to identify these misconfigurations is investigated. A case study is presented to illustrate the direct impact on the end user of conflicting policies amongst xApps along with a potential {AI/ML}-based solution to this problem. This research presents a first analysis of the impact of AI/ML on misconfiguration challenges in {O-RAN}. 
\end{abstract}



\begin{keywords}
Open RAN \sep O-RAN \sep ML \sep 5G \sep 6G \sep Security \sep xApp \sep Misconfiguration
\end{keywords}

\maketitle

\section{Introduction}\label{sec:1}
As 5G evolves, the transition to 6G, which is expected beyond 2030 \cite{Kosei2023Hype}, attempts to reinvent human engagement with digital spaces. Extended reality, networked robots, wireless brain-computer interfaces, holographic telepresence, and e-health with body area networks are among the anticipated uses of 6G~\cite{Siriwardhana2021AI6G}. These applications necessitate support for new capabilities for \ac{eMBB}, \ac{URLLC}, and \ac{mMTC}~\cite{IEEE2019IEEEStandard, Masur2022ArtificialIntelligence}. 
To achieve these goals, major reshaping of existing 5G and 6G architectures is necessary, with a focus on offering flexibility, configurability, and automation.

The \ac{RAN}, a critical and costly component in wireless networks, is part of the innovation in 5G and 6G. This component, which may be considered the most complex part of cellular networks, is undergoing transition through technologies such as \ac{O-RAN}\footnote{\textcolor{black}{The study is based on O-RAN, which refers to the Open RAN architecture defined by the O-RAN ALLIANCE.}}. \ac{O-RAN} has a disaggregated, virtualized, and software-based strategy, linking components via open interfaces and enabling interoperability among vendors~\cite{Polese2023Understanding}. Furthermore, the \ac{AI/ML} integration in O-RAN enables intelligent management of RAN resources, addresses optimisation challenges, and elevates the user experience \cite{Soltani2022Can}. Particularly, \ac{O-RAN} introduces the \acl{RIC} 
 (RIC), which houses third-party applications (rApps and xApps) powered by \ac{AI/ML} that streamline \ac{RAN} operations and manage complexity~\cite{Choongil2024Standardization, Giannopoulos2022Supporting}.

As a result, unlike previous \ac{RAN} technologies, O-RAN has the potential to provide programmability, optimisation, and end-to-end automation in 5G and 6G.  However, realising this potential is dependent on the correct configuration and operation of O-RAN components. Neglecting these factors may result in a variety of misconfiguration difficulties.

Misconfiguration is defined by the \ac{NIST} as \textit{an incorrect or suboptimal configuration of an information system or system component that may lead to vulnerabilities}~\cite{Johnson2011NISTMisconf}. In this respect, misconfiguration allows or induces unintended behaviour, hence impacting a system's security posture \cite{Cook2016NISTMisconf}. Based on these criteria, it could be argued that misconfiguration has a direct and indirect impact. The direct impact is a decrease in system performance, while the indirect impact is an increased vulnerability to security attacks. 

Misconfigurations are more prevalent for the \ac{3GPP} \ac{NG-RAN} than for previous generations such as \ac{UTRAN} and \ac{E-UTRAN}. This increased risk is associated with the introduction of new technologies such as \ac{SDN}, \ac{NFV}, cloud computing, and \ac{AI/ML} in NG-RAN. When combined with the disaggregation and openness envisioned for \ac{O-RAN}, as well as the introduction of third-party applications into the \ac{RIC}, these technologies augment the system's complexity and raise the possibility of misconfiguration. Even minor mistakes in setting up protocols, interfaces, APIs, authentication, and authorization systems might result in new vulnerabilities and security breaches~\cite{Zoure2022Network}.

AI/ML emerges as a possible approach for managing the O-RAN's configuration challenges. It provides automation features for both high-level orchestration and low-level resource optimisation.  Nonetheless, the incorporation of AI/ML with O-RAN presents the possibility of misconfigurations. In this context, a thorough examination of both of these aspects is required in order to comprehend all misconfiguration challenges and engage in discussions about the essential solutions to be adopted.

This paper analyzes misconfiguration concerns in O-RAN, examines the use of AI/ML to identify misconfigurations, and presents a case study that provides insight into the potential consequences of O-RAN system misconfiguration issues. This study draws on a large number of academic publications, white papers from engineering-focused initiatives, and industry and standardisation documents.

\subsection{Motivation}

According to a recent report by Mavenir \cite{Mavenir2021Security}, misconfiguration is the leading cause of cloud-data breaches. Another study by Positive Technologies~ \cite{Positive20195G} found that one in every three successful attacks on 4G networks is caused by faulty equipment configuration. In the context of commercial and open-source software, Zhang et al. \cite{Zhang2021Static} found that misconfiguration accounted for 31\% of server downtime issues, compared to 15\% for software faults. These statistics are relevant to the study of the O-RAN system, which is supported by open-source software and cloud computing. Furthermore, most existing 5G deployments primarily follow the \ac{NSA} approach, indicating a reliance on 4G infrastructure~\cite{Kosei2023Hype, Kosei2023Magic}.

Misconfigurations in O-RAN are critical, yet they have received little attention.  Previous initiatives, including those of research bodies \cite{Mimran2022Security, Madhusanka2023Open, Polese2023Understanding}, telecommunication standardization bodies, such as O-RAN \cite{ORANWG112023ThreatModeling, ORANWG112023SecurityProtocols, ORANWG112023SecurityRequirements}, 3GPP \cite{3GPP2023SecurityAspectsAIML}, cybersecurity agencies, such as ENISA~\cite{ENISA2023CyberSecurityOpenRAN}, engineering-focused initiatives, such as TIP \cite{TIP2023OpenRANSecurity}, and industry documents, such as Mavenir \cite{Mavenir2021Security}, Rimedo Labs~\cite{Bogucka2023RimedoLabsSecurity}, Ericsson \cite{Ericsson2020Security}, Rakuten Symphony~\cite{Rakuten2022ORANSecurity}, VMware~\cite{WMware2021ORANSecurity}, NEC~\cite{NEC2022ORANSecurity}, and others \cite{Maria2022Top, Positive20195G}, have primarily focused on analyzing security threats. These include threat models, security requirements, security procedures, risks, vulnerabilities, and attack vectors. It is worth noting that while these studies provide a comprehensive and informative overview, they do not provide an in-depth examination of the complexities associated with the potential deployment of O-RAN. In contrast to previous research, this article focuses on misconfigurations, which are a major problem for \ac{MNO}s due to their potential to degrade network performance and expose the system to security threats.

Misconfigurations are unavoidable in O-RAN owing to its open nature \cite{Madhusanka2023Open}. O-RAN supports multiple vendors' elements (e.g., \ac{RU}, \ac{DU}, \ac{CU}, and \ac{RIC}), supports different versions of hardware and software (e.g., \ac{E2SM}s), operates across multiple technologies (e.g., multiple \ac{RAT} and \ac{SA} and \ac{NSA} deployments), and facilitates multi-tenancy with different \ac{MNO}s. Furthermore, the system's seamless deployment, integration, and operation depend on the joint efforts of many stakeholders or actors. Managing all of this complexity certainly increases the possibility of misconfigurations.

Human errors, whether made by component developers, integrators, engineers, or operators, are the leading cause of misconfiguration \cite{Mimran2022Security, daSilva2019Armor}. These errors can appear in three forms: slips, which are unintentional errors during the configuration workflow; mistakes, which result from a lack of knowledge in a specific aspect of configuration; and violations, which are intentional errors committed under certain conditions, usually due to a failure to adhere to best practices or rules during peak workload hours \cite{Mushi2019Designing}. Implementing advanced technologies such as \ac{SDN} and \ac{NFV} enhances network configuration accuracy and efficiency. This shift from manual procedures to automated processes reduces mistakes. The fast operation of these automated technologies, however, poses the possibility of increasing error probability.  For example, software-based systems such as \ac{VNF}s may include unnoticed build errors  that have serious implications. Furthermore, even these  advanced tools are operated by people, indicating a susceptibility to errors.

Therefore, it is critical to identify misconfiguration issues within O-RAN and investigate the possibilities of AI/ML to address them in order to improve the efficiency and security of RAN deployments. 

\subsection{Our contributions}

 The contributions are summarized as follows:

\begin{enumerate}

\item We provide an overview of AI/ML deployment options in the O-RAN system and offer detailed examples of actual applications. This highlights areas requiring additional studies and development in the application of AI/ML in O-RAN.

 \item  We provide a detailed analysis of misconfiguration problems in O-RAN, focusing on integration and operation, the use of SDN and NFV, and the use of AI/ML. Extensive examples are provided for each type of misconfiguration to aid understanding of the issues. To the best of our knowledge, this is the first analysis of misconfiguration issues in the context of O-RAN. This analysis reveals both opportunities for novel research solutions and identifies critical issues that must be addressed by network providers in their deployment of \ac{O-RAN}.

\item We provide an analysis of misconfiguration detection approaches and emphasize how AI/ML can be employed for detection. Examples of \ac{KPI}s for each misconfiguration type are also provided. 

\item We present an illustrative example of the impact of conflicting xApps to highlight the potential consequences of O-RAN misconfigurations and the potential of AI/ML to identify them. 

\end{enumerate}

The remainder of this paper is organized as follows. Section~\ref{sec:2}  presents the background of \ac{O-RAN} and the application of \ac{AI/ML} within \ac{O-RAN}. The misconfiguration issues in \ac{O-RAN} are analyzed in Section~\ref{sec:3}. Section~\ref{sec:4} reports metrics and detection approaches for  misconfiguration based on \ac{AI/ML}. This section also introduces the case study of detecting conflicting xApps. Finally, the conclusion and future research are presented in Section~\ref{sec:5}.
 Table \ref{tab1} presents important acronyms and definitions used in this document.
 
\begin{table*}
\caption{\textbf{List of important acronyms and definitions.}}
\begin{tabular}{p{1.5cm} p{6.5cm} p{1.7cm} p{6.2cm} }
\hline
\acs{3GPP} & \acl{3GPP}  & \acs{NG-RAN} & \acl{NG-RAN} \\ 
    \acs{5GC} & \acl{5GC}  & \acs{NDT} & \acl{NDT} \\
    \acs{A1} & \acl{A1}  & \acs{Non-RT} & \acl{Non-RT} \\
    \acs{AI/ML} & \acl{AI/ML}  & \acs{NS} & \acl{NS} \\
    \acs{ANN} & \acl{ANN}  & \acs{NSA} & \acl{NSA} \\
    \acs{CNN} & \acl{CNN}  & \acs{O1} & \acl{O1} \\
    \acs{CU} & \acl{CU}  & \acs{O2} & \acl{O2} \\
    \acs{CU-CP} & \acl{CU-CP}  & \acs{O-eNB} & \acl{O-eNB} \\
    \acs{CU-UP} & \acl{CU-UP}  & \acs{O-RAN} & \acl{O-RAN} \\
    \acs{DNN} & \acl{DNN}  & \acs{O-RAN-SC} & \acl{O-RAN-SC} \\
    \acs{DoS} & \acl{DoS}  &  \acs{PCA} & \acl{PCA} \\
    \acs{DU} & \acl{DU}  & \acs{PRB} & \acl{PRB} \\
    \acs{E1} & \acl{E1}  & \acs{RL} & \acl{RL} \\
    \acs{E2} & \acl{E2}  & \acs{RAN} & \acl{RAN} \\
    \acs{E2SM} & \acl{E2SM}  & \acs{RAT} & \acl{RAT} \\
    \acs{E-UTRAN} & \acl{E-UTRAN}  & \acs{RLF} & \acl{RLF} \\
    \acs{eMBB} & \acl{eMBB}  & \acs{RIC} & \acl{RIC} \\
    \acs{F1} & \acl{F1}  & \acs{RRM} & \acl{RRM} \\
    \acs{FH} & \acl{FH}  &  \acs{RU} & \acl{RU} \\
    \acs{FL} & \acl{FL}  & \acs{SA} & \acl{SA} \\
    \acs{FCAPS} & \acl{FCAPS}  &  \acs{SDN} & \acl{SDN} \\
    \acs{IDS} & \acl{IDS}  &  \acs{SLA} & \acl{SLA} \\
    \acs{KPI} & \acl{KPI}  &  \acs{SMO} & \acl{SMO} \\
    \acs{LSTM} & \acl{LSTM}  & \acs{SON} & \acl{SON} \\
    \acs{MAC} & \acl{MAC}  & \acs{UE} & \acl{UE} \\
    \acs{MDP} & \acl{MDP}  & \acs{URLLC} & \acl{URLLC} \\
    \acs{MNO} & \acl{MNO}  & \acs{UTRAN} & \acl{UTRAN} \\
    \acs{MLB} & \acl{MLB}  & \acs{VNF} & \acl{VNF} \\
    \acs{mMTC} & \acl{mMTC}  & \acs{VM} & \acl{VM} \\
    \acs{MITM} & \acl{MITM}  & \acs{WG} & \acl{WG}\\ 
    \acs{NIST} & \acl{NIST}  & \acs{X2/Xn} & \acl{X2/Xn}\\ 
    \acs{Near-RT} & \acl{Near-RT}  & \acs{xHaul} & \acl{xHaul}\\ 
    \acs{NFV} & \acl{NFV}  &  \acs{Y1} & \acl{Y1}\\
    \acs{NG} & \acl{NG} & &\\
\hline
\end{tabular}
\label{tab1}
\end{table*}

\section{Background} \label{sec:2}

This section describes the architecture of O-RAN as well as the integration of AI/ML into O-RAN.

\subsection{O-RAN architecture}

\begin{figure*}[!t]
\centering
\includegraphics[width=\linewidth]{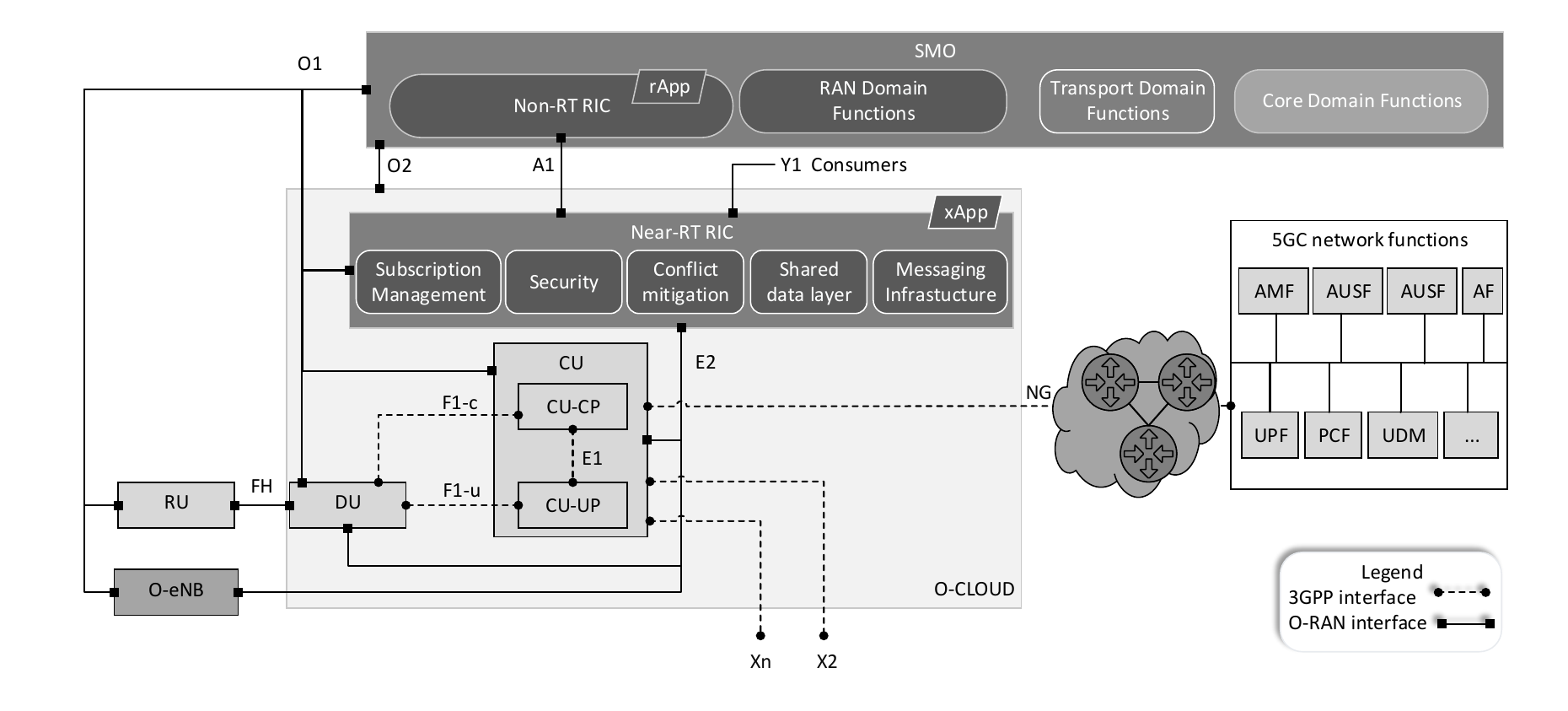}
\caption{O-RAN architecture presented by the O-RAN Alliance and 3GPP \cite{ORANWG12023Architecture} (the RAN is connected to the 5GC through the NG interface).}
\label{fig:o-ran-architecture}
\end{figure*}

Figure \ref{fig:o-ran-architecture} depicts the O-RAN architecture, as defined by the O-RAN Alliance \cite{ORANWG12023Architecture}. Table \ref{tab1} contains the definitions of the components. The \ac{RAN} is divided into three components: the \ac{CU}, \ac{DU}, and \ac{RU}, each of which handles the \ac{NG-RAN} protocol stack in various split configurations \cite{Sirotkin20205G}. The \ac{CU} is subdivided into \ac{CU-CP} and \ac{CU-UP}, which are in charge of \ac{RRM} in the control plane and user plane, respectively. The E1 interface connects the \ac{CU-CP} and \ac{CU-UP}, while the F1 interface connects them to the \ac{DU}. The \ac{DU} and \ac{RU} are linked together by the \ac{FH} interface.

The \ac{RIC} is the central component of the O-RAN architecture. The RIC is divided into two parts: the \ac{Near-RT-RIC} and \ac{Non-RT-RIC} that handle \ac{RAN} resources on millisecond and second scales, respectively. The \ac{Near-RT-RIC} uses xApps (third-party apps) to control the CU, DU, and  RU. The CU/DU/RU are represented as E2 nodes that expose \ac{E2SM}s to the \ac{Near-RT-RIC}. The \ac{E2SM}s  describe RAN functions in an open and standardised manner.  The \ac{Near-RT-RIC} also includes platform services such as subscription management, security, conflict mitigation,  shared data layer, and message infrastructure. Although the O-RAN \ac{WG}3 \cite{ORANWG32023RICARCH} has standardised some of these services, detailed specifications for the majority of them are still pending. In addition to xApps, the functioning of the \ac{Near-RT-RIC} is determined by policies received from the A1 interface and Y1 consumers.

The \ac{Non-RT-RIC}, located inside the \ac{SMO}, is in charge of the long-term objectives in the \ac{RAN}. The \ac{Non-RT-RIC} does this by using rApps, which are third-party applications that build policies to operate xApps over the A1 interface. Furthemore, the \ac{Non-RT-RIC}, in combination with other RAN domain functions in the \ac{SMO}, facilitates RAN domain operation. The \ac{SMO}'s higher level of orchestration allows the development of end-to-end solutions by integrating functions across the RAN, transport, and 5G core domains.

The O1 interface allows communication between the SMO and the RIC, as well as between the SMO and the E2 nodes for \ac{FCAPS}. Furthermore, the O-RAN  connects to the service-based \ac{5GC} via the NG interface, and to other g-NBs and e-NBs via the Xn and X2 interfaces, respectively. The O-RAN architecture also integrates O-eNB, which represents a monolithic RAN deployment, with the capabilities of \ac{E2SM}. Finally, the O2 interface is a critical component of O-RAN, connecting the \ac{SMO} to the cloud platform (O-Cloud).

For the first time, the \ac{3GPP} 5G-Advanced Rel. 18 has standardised the use of AI/ML in the operation of 5G NR~\cite{ORANWG12023Architecture, Polese2023Understanding}. The next section examines the deployment possibilities for \ac{AI/ML} within the O-RAN architecture.

\subsection{AI/ML in O-RAN}
\begin{figure*}
\begin{forest}
forked edges,
for tree = {
    rounded corners, 
    top color=white!5, bottom color=white!30, 
    draw=black, 
    align=center, 
    anchor=children,
    l sep=7mm,
    fork sep = 4mm
            },
before packing = {where n children=3{calign child=2, calign=child edge}{}},
before typesetting nodes={where content={}{coordinate}{}},
where level<=1{line width=1pt}{line width=1pt},
[ML deployment in O-RAN, blur shadow, draw=gray, top color=gray!80, bottom color=gray!30
    [{ Single}, draw=gray, top color=gray!80, bottom color=gray!30
        [RIC function, draw=gray, top color=gray!80, bottom color=gray!30]
        [rApp]
        [xApp]
        [CU or DU]
    ]
    [Distributed, draw=gray, top color=gray!80, bottom color=gray!30
          [Coordinating apps]
            [Model splitting, draw=gray, top color=gray!80, bottom color=gray!30]
          [Model sharing, draw=gray, top color=gray!80, bottom color=gray!30]
          [Federated learning, draw=gray, top color=gray!80, bottom color=gray!30]
          ]
      ]
\end{forest}
    \caption{Deployment options of \ac{AI/ML} models in O-RAN. In gray boxes: Minimally or unexplored areas.}
    \label{fig:deployment_options_AIML_O-RAN}
\end{figure*}
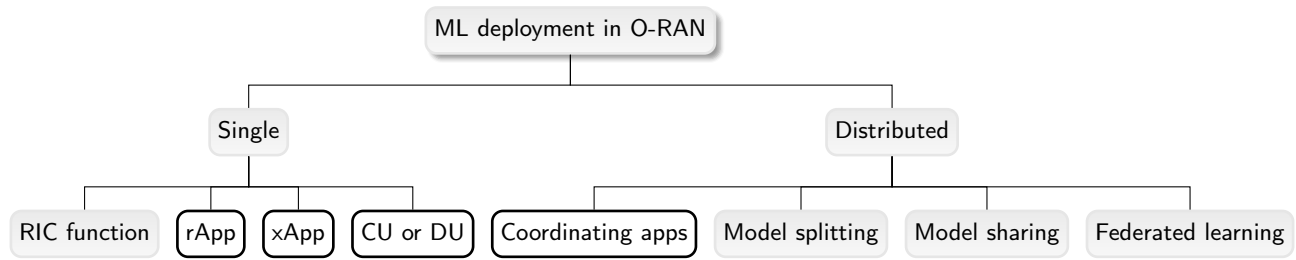

Figure \ref{fig:deployment_options_AIML_O-RAN} shows different options to deploy \ac{AI/ML} models in the O-RAN system according to the \ac{O-RAN} and \ac{3GPP}~\cite{Xingqin2023, 3GPP2023AIMLrequeriments, 3GPP2022EUTRANR}. These deployment options depend on the use case and thus cover different requirements.

In the {\color{black} single} deployment option, the AI/ML model is positioned within a specific component of the O-RAN system. Aside from xApps and rApps, the RIC platform functions, as well as the \ac{CU} and \ac{DU}, can serve as hosts for AI/ML models, handling complex processes. In the O-RAN \ac{Near-RT-RIC} architecture, for example, the conflict mitigation function is integrated into a RIC platform function~\cite{ORANWG32023RICARCH}, which is expected to employ AI/ML-based conflict detection for xApps.

In the distributed deployment option, the AI/ML models are distributed across various O-RAN components. In the coordinated apps, two or more AI/ML-based applications can collaborate to handle integrated challenges, such as employing coordinated solutions based on rApps and xApps. Model splitting divides and assigns the AI/ML model to various linked parts. For example, some layers of an \ac{ANN} might be assigned to the end-device (user equipment), while the remainder layers are placed in a RIC application. Model sharing entails centralising the model in a component with high availability, high processing and storage capabilities, and enabling other O-RAN system parts with lower capabilities to download the model as needed. Finally, the employment of \acl{FL} (FL) models in O-RAN is expected to address distributed applications while protecting user privacy. 

Table \ref{tab2} shows examples of applications that employ AI/ML in O-RAN. The current trend in academia and industry is to investigate { single} deployments, particularly xApps and rApps based on AI/ML for energy saving, load balancing and mobility optimisation, anomaly detection, and \acl{NS} (NS). It is worth noting the collaboration between various industries and academic entities in the creation and testing of xApps and rApps (see Table~\ref{tab2}). The cooperative effort of NetAI and VMware \cite{ORAN2023PlugFest}, where they demonstrated their energy-saving rApp, is an example of such collaboration. The \ac{O-RAN-SC} \cite{ORANSC2023RICAPP} contains a collection of open-source xApps created and validated by a variety of companies, including but not limited to AT\&T, Rimedo Labs, and others. These collaborations provide a substantial contribution to the progress of O-RAN.

\begin{table*}[]
\centering
\caption{Examples of existing xApps or rApps developed by industry, research bodies, and the \ac{O-RAN-SC} for different use cases. The applications can be deployed as { single (Sing)}  or distributed (Dist). Most of these applications use \ac{AI/ML}.}\label{tab2}
\begin{tabular}{p{1.6cm} p{2.3cm} p{5.7cm} p{0.6cm} p{4.4cm}}
 \hline
 \rowcolor[HTML]{C0C0C0}
Use case & Provider & Application details & Dply & \ac{AI/ML} details \\ 
 \hline
  & Rimedo Labs \cite{Rimedo2023ORAN}& rApp switches on/off cells based on user throughput and power consumption.       & { Sing}  & Uses \ac{RL}.            \\ \cline{2-5} 
    & ONF \cite{ONF2023Fyuz}& rApp monitors the load of cell 1 and decides to switch it off. The xApp moves the traffic from cell 1 to cell 2, selected by the rApp. &   Dist & Not specified.            \\ \cline{2-5} 
  & Nokia   \cite{ORAN2023PlugFest}      &    xApp guides gNBs to cover areas served by other gNBs, enabling the shutdown of those cells. & { Sing}   &  Not used.         \\ \cline{2-5} 
  &  Ericsson  \cite{Ericsson2023EnergyrApp}      &     rApp monitors the radio units and network, understands root causes of inefficiencies, and provides recommendations for resolution.    & { Sing}   & \ac{AI/ML} for network clustering and modeling (\ac{AI/ML} model not specified).      \\ \cline{2-5} 
\multirow{-10}{*}{\begin{tabular}[c]{@{}l@{}} Energy \\ saving \end{tabular}}                                 &  Net AI and VMware  \cite{ORAN2023PlugFest}        &  xApp for carrier number forecasting for improved energy efficiency.            & { Sing}   &   Net-AI forecasting engine (\ac{AI/ML} model not specified).        \\ \hline
\rowcolor[HTML]{C0C0C0} &    Orhan et. al.   \cite{Orhan2021ConnectionMgmt}                                  &  xApp for user-cell association and load balancing.     &         { Sing}   & The problem is formulated using graph ANN and solved using DRL.          \\ \cline{2-5} 
\rowcolor[HTML]{C0C0C0}                                        & Rimedo Labs \cite{Rimedo2023ORAN}         &  xApp commands handover operations based on A1 policies.             & { Sing}   & Not used.         \\ \cline{2-5} 
\rowcolor[HTML]{C0C0C0}                              &   O-RAN-SC \cite{ORANSC2023RICAPP}   & xApps for load prediction by CCMC and traffic steering by AT\&T and UTFPR. & { Sing}  &  Not specified.   \\ \cline{2-5} 
\rowcolor[HTML]{C0C0C0} & Lacava et. al. \cite{Lacava2022Programmable}  &  xApp maximizes the UE throughput utility through handover control.   & { Sing}   &  The problem is formulated as a \ac{MDP} and solved using \ac{RL}.   \\ \cline{2-5} 
\rowcolor[HTML]{C0C0C0} & Mahrez et. al. \cite{Mahrez2023Benchmarking}   &  xApp balances the load across cells and optimizes the handover process. It uses xApps for KPI monitoring and anomaly detection.  & { Sing}   &  Uses isolation forest model to detect anomalous UEs.   \\ \cline{2-5} 
\rowcolor[HTML]{C0C0C0} & Ntassah et. al. \cite{Ntassah2023xAppTS}   &  xApp performs UE handovers based on the predicted cell throughput. UE clustering speeds up decision-making.  & { Sing}    & K-means for UE clustering and LSTM for cell throughput prediction.  \\ \cline{2-5} 
\rowcolor[HTML]{C0C0C0} & Kasuluru et. al. \cite{Kasuluru2023ProbabilisticForecasting}   &  xApp forecasts the demand of PRBs of the \ac{CU}. & { Sing}   & Probabilistic forecasting: Transformers, Simple-Feed-Forward, and DeepAR.  \\ \cline{2-5} 
\rowcolor[HTML]{C0C0C0} & Boutiba et. al. \cite{Boutiba2023DynamicTDD}   & xApp for dynamic time duplex division (D-TDD). & { Sing}    &  Deep deterministic policy gradient for optimal TDD configuration based on uplink/downlink demands. \\ \cline{2-5} 
\rowcolor[HTML]{C0C0C0}
\multirow{-22}{*}{\cellcolor[HTML]{C0C0C0}\begin{tabular}[c]{@{}l@{}}Load \\ balancing \\ and \\ mobility \\ optimization\end{tabular}}        & Mavenir  \cite{Mavenir2023xrApps}       &  xApps and rApps for load distribution, \ac{MLB}, \ac{MRO}, coverage and capacity optimization (CCO), automatic neighbor relation (ANR), beam control, smart scheduler, and traffic steering.            & { Sing}   & \ac{AI/ML} utilized without specific details being provided.         \\ \hline
                 &  Ericsson  \cite{Ericsson2023PerformanceDiagnostics}      &   rApp analyzes the RAN to detect and classify cell issues.          & { Sing}    &  \ac{AI/ML} is used to detect anomaly cells, classify coverage, handover or external issues, and correlate each issue to its root cause level.        \\ \cline{2-5} 
\multirow{-3}{*}{\begin{tabular}[c]{@{}l@{}} Anomaly \\ detection \end{tabular}}                     &  Kryszkiewicz et. al. \cite{Kryszkiewicz2023Open}  (Rimedo Labs)      &    xApp  detects jamming attacks based on Reference Signal Received Power (RSRP) and Channel Quality Indicator (CQI) values reported by UEs.          & { Sing}   &   Kolmogorov–Smirnov (KS) is used to detect distribution changes.         \\ \cline{2-5} 
\end{tabular}
\end{table*}

\begin{table*}[]
\centering
\begin{tabular}{p{1.6cm} p{2.3cm} p{5.7cm} p{0.6cm} p{4.4cm}}
 \hline
 \multicolumn{5}{l}{Continuation of Table \ref{tab2}}
 \\
 \rowcolor[HTML]{C0C0C0}
Use case & Provider & Application details & Dply & \ac{AI/ML} details \\ 
 \hline
                 &  Hoffmann et. al. \cite{Hoffmann2023Signaling} (Rimedo Labs)      &    xApp is used to model \ac{KPI} profiles based on the Random Access Channel response. & { Sing}   &     Anomaly detection based on the mean value and standard deviation of the \ac{KPI}.       \\ \cline{2-5} 
                 &  Huang et. al. \cite{Huang2023DevelopingxApps}    &  xApp uses the signal strength stability feature to detect rogue base stations (RBS).   &   Dist & 
The xApp trains the models RF, KNN, and SVM and transfers them to the UE for RBS' detection.         \\ \cline{2-5} 
                 
\multirow{-6}{*}{\begin{tabular}[c]{@{}l@{}} Anomaly \\ detection \end{tabular}}                  &    O-RAN-SC \cite{ORANSC2023RICAPP}        &    Anomaly detection by HCL, \ac{KPI} monitor by Samsung, signaling storm detection by Samsung.            & { Sing}   &       Not specified.   \\ \hline
\rowcolor[HTML]{C0C0C0}  &  Johnson et. al. \cite{Johnson2021NexRAN}   &   NexRAN xApp performs closed-loop RAN slicing control, using \ac{E2SM}s for \ac{KPI} monitoring and \ac{NS}.        & { Sing}   &   Not used.         \\ \cline{2-5} 
\rowcolor[HTML]{C0C0C0}  &  Yeh et. al. \cite{Yeh2024DeepLearning} (Intel)   &  xApp determines the quantity of radio resource for each \ac{NS} and the MAC  schedules and enforces these allocations.      & { Sing}   &   LSTM, temporal CNN, and Seq2Seq for traffic load prediction.          \\ \cline{2-5} 
\rowcolor[HTML]{C0C0C0}  &  Mallu et. al. \cite{Mallu2023AIMLNS}   & rApp sets policies that regulate xApp behaviour while slicing. These rules govern how xApp manages RAN resources.        &   Dist &   ML for policy selection.          \\ \cline{2-5} 
\rowcolor[HTML]{C0C0C0}  &  Wiebusch et. al. \cite{Wiebusch2023TowardsO6G}   &  xApp predicts uplink resource requirements for UEs.        & { Sing}   &   LSTM for UE's payload prediction.          \\ \cline{2-5} 
\rowcolor[HTML]{C0C0C0}  &   \textcolor{black}{Tsampazi et. al. \cite{Tsampazi2023Comparative}}   &  \textcolor{black}{xApp allocates the PRBs for each slice and decides which MAC scheduling is used per slice.}        & \textcolor{black}{ Sing}   &  \textcolor{black}{Deep RL is used for slicing and scheduling (12 DRL designs are tested.)}          \\ \cline{2-5} 
\rowcolor[HTML]{C0C0C0}\multirow{-12}{*}{\begin{tabular}[c]{@{}l@{}} Network \\ slicing \end{tabular}}   &   Zhang et. al. \cite{Zhang2022FederatedNS}      &  Power control xApp and slice-based resource allocation xApp are coordinated to  optimize the use of resources in O-RAN.           &   Dist &  The two xApps use \ac{MDP} and \ac{RL}, and are coordinated with \ac{FL}.   \\ \hline  
\end{tabular}
\end{table*}

Coordinating apps based on AI/ML are emerging for distributed solutions, such as the energy-saving platform tested by Rimedo Labs and ONF \cite{ONF2023Fyuz}, where a traffic steering xApp and an intelligent cell on/off rApp collaborate to optimise \ac{RAN} energy consumption while maintaining service quality. Another study in \cite{Mallu2023AIMLNS} showcased an rApp that creates policies to control the behaviour of a RAN slicing xApp. The rApp selects resource allocation policies in the RAN slices using AI/ML, whereas the xApp implements such policies in near real-time. 

Furthermore, model sharing has recently been investigated in \cite{Huang2023DevelopingxApps}. In this study, an xApp situated in the \ac{Near-RT-RIC} trains an ML model to detect rogue base stations (RBS). The model is then transferred to the UEs, which use it to detect RBSs. The benefits of this technique include real-time detection, lower computational burden on the UE, and the fact that the models of all UEs in the RAN may be updated at the same time. More research is needed to deploy AI/ML in RIC platform functions as well as for model splitting, model sharing, and FL (see Figure \ref{fig:deployment_options_AIML_O-RAN}).

As illustrated, the deployment of \ac{AI/ML} in O-RAN has the potential to autonomously and efficiently manage \ac{RAN} resources. Nonetheless, their implementation poses potential configuration challenges, which are included in the analysis presented in the next section (in particular, Section \ref{sec:3C}).

\section{Misconfiguration problems in O-RAN} \label{sec:3}
This section explores O-RAN misconfiguration issues in terms of integration and operation, enabling technologies, and \ac{AI/ML}. Table \ref{tab3} presents instances of these misconfiguration issues. The table also includes the impacted components and threats linked to each scenario. A misconfiguration problem, according to its definition, can influence either directly, impacting O-RAN performance, or indirectly, presenting a risk of greater susceptibility to threats to security inside O-RAN. These threats are also depicted in the table.

\begin{table*}[]
\centering
\caption{Misconfiguration problems in O-RAN: Examples of misconfiguration, impacted components, and potential direct (performance) and indirect (security) threats are shown. The example ID is provided for reference in the association with detection approaches in Table \ref{tab4}.}\label{tab3}
\begin{tabular}{p{0.8cm}p{1.7cm}p{5.3cm}p{4.1cm}p{3.3cm}}
\hline
 \rowcolor[HTML]{C0C0C0} 
Area  & Aspect   & (ID): Example of misconfigurations &  Impacted components  & Potential threats\\
\hline
&     & (E1): Enabled default ports, services, accounts, and privileges   \cite{ORANWG112023ThreatModeling}.  & Non-RT RIC, Near-RT RIC, CU, DU, RU.  &  Security: intruders. \\ \cline{3-5} 
& & (E2): Lack of conformance or interoperability with standard procedures (e.g., xApp registration).  & Near-RT RIC, Non-RT RIC, xApp, rApps, O2, O1, E1, F1, A1, E2.  & Performance: outages. \\ \cline{3-5} 
& & (E3):  xApps access data from the E2SMs beyond what is strictly necessary. & xApp, CU, DU.  & Performance: monitoring  overhead.  Security: data exposure.  \\ \cline{3-5} 
& & (E4):  Conflicting IP configuration of end-points. & Near-RT RIC, Non-RT RIC, xApp, rApps, CU, DU, RU.  & Performance: outages.   \\ \cline{3-5} 
 & \multirow{-9}{*}{Integration} & \begin{tabular}[c]{@{}l@{}} (E5): Utilizing outdated E2SMs. \end{tabular}  & xApp, CU, DU. &  Performance: outages.  Security: node exposure.  \\
\cline{2-5}  
 & & (E6): Disabled or improper configuration of security protocols (e.g., SSH) to protect reference points~\cite{Mavenir2021Security, ORANWG112023SecurityProtocols}. & A1, O1, O2, E2, F1, E1. &   Security: intruders.  \\  \cline{3-5} 
 & & (E7): Disabled or improper configuration of mutual authentication of endpoints~\cite{Fijitsu2022ABrief}.  & Near-RT RIC, Non-RT RIC, CU, DU, RU. &  Security: rogue endpoints.  \\ \cline{3-5} 
 & & (E8): Lack of failover for endpoint crashes.  & Near-RT RIC, Non-RT RIC, CU, DU, RU. &  Performance: outages.  \\ \cline{3-5} 
  & &  (E9): Sub-optimal balance between  security and CPU  utilization in E2 encryption \cite{Groen2023Implementing}.  &  CU, DU, RU. &  Performance: high CPU usage.  \\ \cline{3-5} 
 &  \multirow{-9}{*}{\begin{tabular}[c]{@{}l@{}} Security \\ function \end{tabular}}    &  (E10): Sub-optimal equilibrium between FH protection and performance~\cite{Dik2021Transport, Lipps2023Designing, Harrilal2023Performance}. & FH. & Performance: high delay and low throughput. \\
 \cline{2-5}
 & & (E11): Previous xApp not uninstalled before new installation. & xApp, CU, DU. & Performance: instability.  \\ \cline{3-5} 
 & &  (E12):  Sub-optimal rule generation. &  Non-RT RIC, Near-RT RIC, xApp,  rApp, CU, DU.   &  Performance: resource wastage.  \\ \cline{3-5} 
 &   & (E13):   A1 policies demand more resources than are available.  & Near-RT RIC, xApp, CU, DU. &  Performance: resource depletion.  \\ \cline{3-5}  
 &   &  (E14):   Meeting E2 policies involves the demand for high energy usage by E2 nodes~\cite{Porambage2023Xcaret}.  &  xApp, CU, DU. &  Performance: energy wastage.  \\ \cline{3-5}  
\multirow{-36}{*}{I\&O} & \multirow{-9}{*}{\begin{tabular}[c]{@{}l@{}} Conflicting \\ policies \end{tabular}} &  (E15): Conflicting access to radio resources by xApps. & xApp, CU, DU. & Performance:  instability. \\ \hline
 & & (E16): Malformed packets \cite{Neves2021Dynamic}. & xHaul &   Performance: packet re-transmission.        \\ \cline{3-5}  
&                   & (E17): Unsynchronized controller instances \cite{Wang2019FlowTracer}. & xHaul & Performance: instability.       \\ \cline{3-5} 
&                   & (E18): Sub-optimal controller placement \cite{Weiwei2024MLbased}. & xHaul &  Performance: high latency, low reliability, energy  wastage.       \\ \cline{3-5} 
&                   & (E19): Inconsistent directives between the controller and stateful network devices. & xHaul & Performance: instability.       \\ \cline{3-5} 
   &   \multirow{-8}{*}{SDN}    & (E20): Violation of firewall application \cite{Saied2020AComprehensive}. & xHaul &  Security: DoS, port scanning.                  \\  \cline{2-5}
&                   &   (E21): Sub-optimal initial resource assignment during image creation \cite{Mekki2022Microservices}.   &  Non-RT RIC, Near-RT RIC, xApp, rApp, CU, DU.  &   Performance: resource wastage, container halting.   \\   \cline{3-5} 
\multirow{-8}{*}{ \begin{tabular}[c]{@{}l@{}} SDN \\ \& \\ NFV \end{tabular} } & \multirow{-3}{*}{NFV}                         &  (E22): Sub-optimal service migration \cite{Huff2021RFT, Ramanathan2021Live}.  &   Non-RT RIC, Near-RT RIC, xApp, rApp, CU, DU.  &    Performance: service downtime.          \\ \cline{3-5}  
\end{tabular}
\end{table*}

\begin{table*}[]
\centering
\begin{tabular}{p{0.8cm}p{1.7cm}p{5.3cm}p{4.1cm}p{3.3cm}}
\hline
\multicolumn{5}{l}{Continuation of Table \ref{tab3}}
\\
\hline
 \rowcolor[HTML]{C0C0C0} 
Area  & Aspect   & (ID): Example of misconfigurations &  Impacted components  & Potential threats\\
\hline
&                        &  (E23): Excessive fragmentation of vDU functions across numerous microservices \cite{Sirotkin20205G}.   & DU.  &    Performance: high latency, intensive inter-service comms.   \\ \cline{3-5} 
&                        & (E24): Incorrect timing/sync between vDU and RU  (PTP) \cite{Sirotkin20205G}.  & DU, RU.  &   Performance: low  reliability.    \\ \cline{3-5} 
\multirow{-5}{*}{ \begin{tabular}[c]{@{}l@{}} SDN \\ \& \\ NFV \end{tabular} } & \multirow{-5}{*}{NFV}                                                                                     & (E25): Lack of VM/container isolation  \cite{Oqaily2022MLFM}. &  Non-RT RIC, Near-RT RIC, xApp, rApp, CU, DU. &   Performance: inconsistency.  Security: intruders. \\ \hline 
&     & (E26): Sub-optimal granularity for data collection \cite{Wen2022AFineGrained}: Reliability vs. overhead vs. privacy. & A1, E2, CU, DU, xApp, rApp. &    Performance: unreliability. Security: data exposure.          \\ \cline{3-5} 
&     &  (E27): Unreliable AI/ML model sharing.  & rApp, xApps, dApp, E2, A1. & Performance: high end-to-end delay, loss of model data.            \\ \cline{3-5} 
&     & (E28): Misplacement of AI/ML model object \cite{DOro2022OrchestRAN}. & rApp, xApps, dApp. & Performance: high end-to-end delay.            \\ \cline{3-5} 
& \multirow{-8}{*}{\begin{tabular}[c]{@{}l@{}} Performance \\ and \\ reliability \end{tabular}}    & (E29): Implicit conflicts between AI/ML decisions. & rApp, xApp, dApp, CU, DU. & Performance: instability.            \\ \cline{2-5} 
 &    & (E30): Sub-optimal protection of training data:  encryption vs. reliability.  & xApp, rApp, dApp. & Performance: high end-to-end delay.  Security:  poisoning attacks.        \\ \cline{3-5} 
 & \multirow{-3}{*}{\begin{tabular}[c]{@{}l@{}}  Model\\ protection \end{tabular}}      & (E31): Lack or improper anonymization of UE information. & E2, A1. &  Security: data exposure.     \\ \cline{2-5} 
&  & (E32): Use of DNN where not needed. & xApp, rApp, dApp. & Performance: low accuracy, resource wastage. Security: adversarial attacks. \\ \cline{3-5}       
\multirow{-20}{*}{AI/ML}    &  \multirow{-3}{*}{Explainability} & (E33): Too complex design of AI/ML model. & xApp, rApp, dApp. &  Performance: low  accuracy. Security: lack of trustworthiness. \\ \hline
\end{tabular}
\end{table*}

\subsection{Integration and operation}
As the O-RAN has numerous manufacturers, \ac{RAT}s (e.g., WiFi and \ac{NR}), \ac{UE}s (e.g., vehicles and \ac{IoT}), software versions (e.g., \ac{E2SM}s), applications (e.g., \ac{eMBB} and \ac{URLLC}), and so on, it is exceedingly difficult to integrate and operate.  The misconfiguration issues that may arise in this context are discussed below.

    \subsubsection{Integration}   
    
    In an O-RAN, the lack of developed standard procedures might lead to uneven deployment. For example, noncompliance with typical xApp discovery, registration, and subscription processes in the \ac{Near-RT-RIC} will impact automated xApp deployment. Furthermore, because current O-RAN apps coexist in 5G and 4G (NR and E-ULTRA) in \ac{SA} and \ac{NSA} deployments, this integration might cause several setup issues. For example, the complex process of integrating LTE and 5G into \ac{NSA} installations necessitates a careful setup and orchestration. This can help avoid mistakes that might jeopardize the overall system performance, such as bottlenecks and resource underutilization \cite{tele2023RiskAware}.  

Inadequately built architectures; the use of unneeded or insecure parts (ports, services, accounts, privileges), functions, protocols, and components; and dependence on default configurations are other examples of integration issues. These misconfigurations not only expose the system to prospective attackers but also degrade system performance \cite{ORANWG112023ThreatModeling}.

    \subsubsection{Security function}
    
    Three components are required to enable effective O-RAN protection: (i) protecting communication at all interfaces, (ii) guaranteeing the trust-based authentication of communicating endpoints, and (iii) leveraging trusted certificate authorities for identity provisioning \cite{Mavenir2021Security}. The \ac{3GPP} and \ac{O-RAN} Alliance released security assurance standards for the O-RAN interfaces, including backhaul, midhaul (F1), \ac{FH}, O1, E2, A1, O2, E1, and Xn \cite{ORANWG112023SecurityProtocols}. These requirements strive to reduce the threat surfaces to provide O-RAN confidentiality, integrity, and replay protection.  In particular, a set of well-proven security protocols, such as SSHv2, \ac{TLS}, DTLS, \ac{IPsec}, and \ac{MACsec}, was chosen.
    
    Nonetheless, the complexities of security protocols, including several sophisticated setups and details, render these protocols vulnerable to misconfigurations. To facilitate the deployment of these protocols, their settings can be incorporated into open-source SSL/TLS libraries \cite{Mimran2022Security}. However, improper use of these libraries exposes the network to the introduction of rogue \ac{RU}s, \ac{DU}s, \ac{CU}s, or \ac{RIC}s. Rogue elements can lead to \ac{MITM} attacks that eavesdrop, change, stop, or delay messages in both the control and user planes \cite{Madhusanka2023Open,  Harrilal2023Performance}.

Furthermore, adding strong protection measures for interfaces with strict timing constraints might reduce the O-RAN performance. The security of the \ac{FH} interface is an example of this difficulty. An optimal security protocol option — \ac{TLS}, \ac{IPsec}, or \ac{MACsec} — must consider the overhead associated with bandwidth and latency \cite{Harrilal2023Performance}.

Finally, sensitive data in the ORAN system should also be protected. These data include the following: (i) data from system functions, such as logging messages, configuration file exports, CLI, or GUI configurations; (ii) authentication data, such as PINs, passwords, cookies, and cryptographic keys; and (ii) data from system elements, such as UE information, RAN topology information, and ML databases, which contain critical information from the system \cite{3GPP2023SecurityAssurance}.

    \subsubsection{Conflicting policies}
    
The total automation of the O-RAN architecture in 5G requires global orchestrators such as the \ac{SMO}, as well as local orchestrators such as the \ac{RIC}. Through a human-machine interface, these orchestrators allow operators to communicate their objectives in a high-level language. These intents are subsequently turned into policies that regulate and run various O-RAN system components.

Many misconfigurations might occur when administering policies in O-RAN. When converting intents into low-level rules for operating system components, for example, the quantity of rules created may exceed the system's resources. Furthermore, the time necessary for rule creation, enforcement, and verification may surpass the performance requirements \cite{Pan2023Misconfiguration, Kermabon2022Prospec}.

When several actors seek to manage a function, policy violations become a big challenge. This situation is demonstrated in the O-RAN system by the functioning of an xApp, which receives policies via the A1, O1, and Y1 interfaces. In such cases, failures might develop due to the interplay of different components introducing contradictory policies \cite{daSilva2019Armor}. When chaining pieces with diverse functions, each with a unique configuration, conflicting rules represent an increased risk. This complication impacts policy communication, potentially resulting in the development of duplicate, shadowed, correlated, or nested rules coming from different intents \cite{Li2018Rule}. 

The multivendor environment causes conflicts between xApps and rApps in O-RAN. Consider the xApps for conflict power allocation and radio resource allocation~\cite{Zhang2022Team}. In this case, the power allocation xApp may assign a high transmission power to one resource block, while the radio resource allocation xApp assigns this resource block to a user with a low traffic load. This disagreement will waste limited bandwidth and increase power usage. Additional conflicts may develop as a result of \ac{RRM} choices made by O-gNB or O-eNB nodes and xApps, potentially creating network instability.

Finally, efficient policy management requires using as few resources as possible inside the O-RAN system, such as containers~\cite{Kermabon2022Prospec}, RUs, and energy usage~\cite{Porambage2023Xcaret}.

\subsection{Enabling Technologies: \ac{SDN} and \ac{NFV}}

O-RAN relies heavily on \ac{SDN} and \ac{NFV} for programmability and flexibility. However, as analyzed below, they also carry the risk of misconfiguration.

\subsubsection{\ac{SDN}}
This technology enables configurable data planes in xHaul networks, as required by end-to-end 5G and 6G systems. Several misconfiguration issues might arise during the functioning of these networks. The integrity of data packets, for example, can be altered by network elements and controlling programmes throughout the transmission process. This effect may manifest as alterations to the packet header, such as changing the VLAN value, resulting in re-transmission events and packet loss. Other data transmission breaches include way-pointing violations, in which packet routes differ from the anticipated device sequence, and traffic locality violations, in which packets must stay inside a defined area \cite{Neves2021Dynamic}.

Emerging data plane programmability technologies, such as P4, hold the potential to increase operational flexibility. However, combining P4 with network controllers presents substantial setup issues, such as selecting which operations are offloaded to P4. Furthermore, independent decisions made by P4 devices may result in discrepancies with the controller, resulting in network instability.

  Flow-based network management, made possible by software-based controllers, is critical in 5G operations. Nonetheless, misconfigurations caused by several coexisting applications in the controller may result in high-level forwarding policies that the data plane cannot follow \cite{ Wang2019FlowTracer}. Furthermore, unsynchronized controller instances and uncontrolled network device failures (e.g., switch port failure) might interrupt flow trajectories. These flaws may cause traffic to be dropped or sent via inappropriate paths.

Finally, firewall applications are among the most significant applications in the programmable data plane. A significant challenge in this domain is ensuring continuous compliance of the data plane with the security policy deployed in the firewall application. Inadequate or incorrect network policy, controller software, and packet trajectory verification might result in partial or total firewall application violations \cite{Saied2020AComprehensive}.

\subsubsection{\ac{NFV}}
Every element inside the O-RAN architecture has the possibility of virtualization using technologies such as virtual machines and containers. This technique improves the \ac{RAN} operations' flexibility and scalability. However, it increases the possibility of misconfigurations.

The initial setup of resources for a virtual component is critical. To achieve optimal service performance, the operator must establish the proper CPU and memory allocation. When the workload exceeds capacity, insufficient resources may result in service failure or decreased performance \cite{ Mekki2022Microservices}. Allocating more resources than necessary for an application, on the other hand, leads to resource waste, which raises deployment costs. 

While it is true that virtualized pieces may be scaled, these operations may cause further configuration problems and disrupt service continuity. For example, virtual service replication and migration are used to solve heavy workload scenarios or in the case of a breakdown. Due to the slow replication process or message rerouting, system state inconsistencies may occur in replication, resulting in inconsistent management between the original and replica services \cite{Huff2021RFT}. Migration, on the other hand, presents issues such as service recovery time and probable data loss \cite{Ramanathan2021Live}. 

Furthermore, the increased complexity of controlling and orchestrating many virtual functions increases the potential for misconfiguration, such as insufficient network isolation between separate network functions \cite{Oqaily2022MLFM}. Similarly, configuration inconsistencies may arise, such as when a virtual firewall defined at the tenant level is possibly circumvented at the underlying cloud infrastructure level. Inconsistencies of \ac{VNF}s might reduce system performance and expose services or infrastructure to threats to security \cite{Lakshmanan2019Modeling, Kotulski2018Towards}.

\subsection{AI/ML} \label{sec:3C}
Given that \ac{AI/ML} is a primary driver of O-RAN advancements and its implementation in O-RAN has already begun (see Table \ref{tab2}) investigating AI/ML misconfiguration issues is critical. In the following, they have been recognised in terms of performance and reliability, model protection, and explainability.

\subsubsection{Performance and reliability}
Misconfigurations throughout the life cycle of \ac{AI/ML} applications can have a negative influence on their reliability and performance. In data collection, for example, improperly setting the data resolution within the system for monitoring leads to problems like: (i) insufficient granularity resulting in inefficient controls, such as failure to identify events; (ii) unnecessary high resolution in the monitoring system resulting in system overhead, such as E2 channel saturation; and (iii) the possible disclosure of sensitive data, such as UE-related details. 

O-RAN suffers from an absence of data monitoring frameworks specialised to \ac{AI/ML} applications~\cite{Wen2022AFineGrained}. For example, the current version of \ac{E2SM} KPM.v3 (O-RAN \ac{WG}3) provides detailed metrics for the \ac{RAN} system that are aligned with 4G LTE (3GPP). These metrics, however, may not fulfil the criteria for particular security solutions, such as strong DoS detectors, which require information at either the packet or flow level.

More proactive E2 nodes can help respond to outages or satisfy the low-latency needs of 6G networks. Integrating pre-processing algorithms, such as \ac{PCA} and auto-encoding, in E2 nodes might help minimise information transfer between these nodes and xApps as well as simplify the \ac{AI/ML} model structure. Similarly, data augmentation approaches, such as generative adversarial learning, may be useful in supplementing data-hungry applications or in cases when O-RAN system samples are insufficient or unbalanced~\cite{Hughes2019Generative, Yea2023AModel, Abdelaty2021GADoT}.

Another misconfiguration issue is the lack of protection of user privacy in data collection. As the \ac{RAN} processes information from all \ac{UE}s, their data privacy must be secured against \ac{AI/ML} activities, which are handled by third-party xApps or rApps in \ac{O-RAN}~\cite{3GPP2022EUTRANR, 3GPP2023SecurityAspectsAIML}.  Details such as \ac{UE} position and trajectory forecasts are examples of privacy-sensitive information. Neglecting data privacy issues exposes the system to the possibility of data leaks, which can result in legal ramifications, monetary fines, loss of customer trust, and harm to the reputation of the entities involved.

In terms of \ac{AI/ML} model performance in O-RAN, if the models fail to achieve the basic requirements given by the use cases, including factors such as accuracy, model size, convergence time, and prediction time, they become unreliable, \cite{Doriguzzi2020lLucid, DOro2022OrchestRAN}. \ac{AI/ML} models may be deployed at multiple locations (see Figure \ref{fig:deployment_options_AIML_O-RAN}) depending on the use case, such as xApps, rApps, and dApps (those deployed at \ac{DU}s and \ac{CU}s~\cite{DOro2022OrchestRAN}). However, if the models are misplaced, they might cause unacceptable delays in the target application's end-to-end control loops, resulting in a decline in system performance rather than a benefit.

For distributed deployments of \ac{AI/ML} (see Figure \ref{fig:deployment_options_AIML_O-RAN}), the O-RAN system must guarantee communication reliability of no less than 99.999\% to support the exchange of data and model parameters and to enable communication across modules or partitions of the models \cite{3GPP2023AIMLrequeriments}. Furthermore, it is critical to provide consistent data feeding as well as the availability of storage and processing resources for \ac{AI/ML} models as and when they are required. Failure in certain arrangements might lead to unreliable model output.

Finally, \ac{AI/ML} model decisions may clash with other functionalities inside the O-RAN system. In particular, the incorporation of \ac{AI/ML} in O-RAN has the potential to generate very complex conflicts, namely implicit conflicts. These conflicts may cause delayed reactions inside the system, making identification a difficult task.

\subsubsection{Model protection}
Adversarial attacks against AI/ML models, including data poisoning, evasion attacks, and API-based attacks, have been investigated in recent years \cite{ENISA2021Securing, Soltani2022Can}. Data poisoning attacks affect the AI/ML model training phase, causing the model to learn incorrectly. Data injection, data manipulation (labels, features, and learning parameters), and logic corruption are examples of these attacks. Evasion attacks, such as  \ac{FGSM} and \ac{PGD}, target the model inference phase. Model extraction, model inversion, and membership inference are examples of API-based attacks that take advantage of the exposure of the \ac{AI/ML} front-end.

 Notably, adversarial attacks have greater entrance hurdles in a monolithic and single-vendor \ac{RAN} architecture since they often lack access to the \ac{AI/ML} models for most applications~\cite{Apruzzese2022Wild}. However, the entrance barriers to such attacks are significantly decreased in the O-RAN system, where the components are disaggregated and third-party suppliers of hardware and software are included~\cite{Benzaid2020AI, Sliwa2022Security, Soltani2022Can}.

Both 3GPP \cite{3GPP2023SecurityAspectsAIML} and O-RAN Alliance \cite{ORANWG112023ThreatModeling} have conducted studies to better understand the risks connected with the usage of AI/ML models. Three threat models against the AI/ML system were found: (1) poisoning attacks, (2) modifying the ML model, and (3) transfer learning attacks~\cite{ORANWG112023ThreatModeling}. The lack or misconfiguration of protection for AI/ML models, as well as the use of public datasets to train the models, are the prevalent flaws across these threat models. 

Recent efforts in \cite{Usama2021Examining, Davaslioglu2019Trojan, Erpek2019Deep, Omara2022Adversarial} have demonstrated the possibility of adversarial attacks on O-RAN operations. However, they are confined to analysing public datasets or employing minimalist testbeds, raising the question: Is there still a risk of adversarial attacks if the security functions have been appropriately established, i.e., safeguarding the communication interfaces and guaranteeing adequate authorization and authentication to access the AI/ML model and data? At first glance, correct security function configuration may avoid poisoning and API-based attacks.  Implementing encryption and decryption methods for training databases at the \ac{Near-RT-RIC}, for example, can function as a measure to protect against data contamination by malicious xApps adopting adversarial approaches, such as \ac{FGSM} and \ac{PGD}~\cite{Naik2023ExperimentalStudy}. However, the question is whether these encryption and decryption operations can be implemented in \ac{O-RAN} without severely influencing the performance of \ac{AI/ML} applications.

In cases of evasion attempts, the persistence of the attacks can be seen regardless of whether the security protections inside the \ac{O-RAN} architecture are appropriately implemented~\cite{Apruzzese2022Wild}. This is because, even with minimal knowledge of \ac{AI/ML} processes in the \ac{RAN}, \ac{UE}s can operate as adversarial agents, impacting the performance of different applications, e.g., automated modulation categorization and forecasting the Channel Quality Indicator (CQI) \cite{Apruzzese2022Wild}.

\subsubsection{Explainability} 
As AI/ML finds use within different areas of O-RAN across 5G and 6G, the lack of explainability within these models may cause significant hesitation, especially when using them in safety-critical use cases such as transportation automation (e.g., trains and \ac{UAV}s), vital infrastructure operation (e.g., water and nuclear energy), healthcare, and human-machine brain interfaces~\cite{Guo2022ExplainableAI6G, Nwakanma2023XAIForIDS}. 
This is especially important when employing \ac{DNN}s, which are data-driven models able to surpass standard mathematical or probabilistic models. Yet, such \ac{DNN}s operate as complex black box models, making it challenging to explain the decisions they make to human specialists, considering the underlying data support and causal logic. 

Poorly designed DNN solutions can exacerbate the explainability problem in O-RAN. For example, when a system's mathematical model is well-established, the employment of \ac{DNN} becomes superfluous. In these circumstances, traditional statistical or signal processing approaches may outperform DNNs. Incorporating DNNs in such settings not only reduces performance and increases vulnerability to adversarial attacks, but it also lacks the critical feature of explainability~\cite{Choongil2024Standardization, Guo2022ExplainableAI6G}.

Furthermore, the adoption of a sophisticated \ac{DNN} architecture with an excessive number of parameters and layers, the employment of complex activation functions, and the lack of preprocessing procedures for input features all contribute to \ac{DNN}s' increased complexity. Such complicated \ac{DNN} architectures are unneeded in many cases. This unnecessary complexity not only raises the processing needs for \ac{DNN} decision-making, but it also increases the danger of model overfitting. It is crucial to highlight that certain data-driven models are intrinsically explainable, such as rule-based models, linear models, Bayesian inference, and decision trees. Depending on the application, these solutions may efficiently replace sophisticated \ac{DNN}s with negligible performance loss.

\subsection{Summary \textcolor{black}{and insights}}

Many of the misconfiguration issues presented in Table~\ref{tab3} have been seen in prior systems that incorporated SDN and NFV. Nonetheless, poor setups of these elements become ever more important in the RAN ecosystem, where resources are scarce and expensive. Furthermore, because the application of AI/ML in RAN is new, its effective integration and operation have the potential to cause significant misconfiguration issues, as seen in Table~\ref{tab3}. 

\textcolor{black}{In addition, the misconfigurations examined in this section may arise at different stages during the implementation lifecycle of the O-RAN system. Recall that misconfigurations are allowed or induced unintended behaviours~\cite{Cook2016NISTMisconf}, as described in Section \ref{sec:1}. For instance, despite a system operator being provided with the standard specifications and industry best practices for securing O-RAN system interfaces (see (E6) in Table \ref{tab3}), the complexity of the system may result in the operator making inadvertent errors. Seemingly minor misconfigurations can lead to significant security impact, such as system intrusions.}

\textcolor{black}{Other misconfiguration problems may be harder to avoid  during the early stages of the lifecycle, such as conflicting xApps (see (E2) in Table \ref{tab3}) in a multi-vendor ecosystem. In such cases, the applications may be correctly developed and integrated into the system,  but their combined configurations may cause conflicts during operation.}

\textcolor{black}{The optimal approach to deal with misconfigurations is to prevent them by adhering to standards, best practices, and employing rigorous verification procedures. Nevertheless, as previously stated, misconfigurations cannot be entirely eradicated, necessitating the consideration of detection strategies, as detailed in the following section.}

\section{AI/ML for misconfiguration detection} \label{sec:4}

\textcolor{black}{This section first describes the problem of misconfiguration detection. Then, it provides an overview of various misconfiguration techniques, emphasizing the role of AI/ML in enhancing detection efficacy.}

\subsection{\textcolor{black}{The problem of misconfiguration detection}}

\textcolor{black}{
In terms of detection, the primary challenge lies in identifying the origin of misconfigurations. For example, if outages occur within our system, various factors could be responsible. One cause might be an xApp failing to adhere to standard procedures when interacting with the E2SMs (see (E2) in Table \ref{tab3}), resulting in its malfunction within the system. Alternatively, conflicts between the xApp and internal RIC functions or other xApps could also lead to outages (see (E15) in Table \ref{tab3}). In this scenario, monitoring the system (using logs, configuration files, KPIs, network packets, etc) or using representation models of the O-RAN system can help locate the problem. Also, it is essential to acknowledge that varied monitoring/detection methods are necessary depending on the specific types of misconfigurations.}

\textcolor{black}{In the next section, several approaches for detecting misconfigurations are examined. Given the intricacies of the O-RAN system, manual or partially automated detection methods might not be sufficient. Therefore, we highlight the benefits of leveraging AI/ML techniques to further improve misconfiguration detection performance.}

\subsection{\textcolor{black}{AI/ML-assisted detection approaches}}
This section explores the potential of \ac{AI/ML} to detect misconfiguration problems in \ac{O-RAN}. In this context, Table \ref{tab4} shows instances of AI/ML-based misconfiguration detection approaches covering various misconfiguration challenges presented in Table \ref{tab3}. Table \ref{tab4} also displays the KPIs for each misconfiguration problem. The misconfiguration detection approaches are described below.

\begin{table*}[]
\centering
\caption{Examples of detection approaches, KPIs, and use of AI/ML for different misconfiguration types in O-RAN.}\label{tab4}
\begin{tabular}{p{2.5cm}p{1.5cm}p{3.5cm}p{4.4cm}p{3.3cm}}
\hline
\rowcolor[HTML]{C0C0C0} 
Misconfiguration ID  & Detection approach & Description &  KPIs &  Use of AI/ML \\ \hline   I\&O-(E1)  & Active monitoring & Port scanning and service scanning.  & Number of open ports,  default accounts, and default passwords. & Data analytic.  \\ 
I\&O-(E2)  &  Passive monitoring & Sniffing of packets for analysis. & Number and type of procedure violations. & Data analytic. \\
I\&O-(E6)  & Active monitoring & Security protocol verification (e.g., SSH~\cite{IETF2012SSHVerification}). & Number and type of check failures. & Data analytic.    \\ 
I\&O-(E9) & Active monitoring & Packet injection and metric collection. & Round trip time, processing delay, transmission delay, throughput, and CPU utilization.   & Anomaly detection.\\
I\&O-(E10) &  Active monitoring & Packet injection and metric collection. & End-to-end latency and throughout.& Anomaly detection.\\
I\&O-(E12)   &  Formal verification  & Model of minimal interval set~\cite{Pan2023Misconfiguration}.  &   Size of generated rule sets, number  of redundant rules, i.e., correlated, shadowing, and imbrication \cite{Li2018Rule}.  &   Learning of the  representation and anomaly detection.\\
I\&O-(E15) & NDT & Creation and testing of risk scenarios. & Number and types of conflicting access to resources.  &  Creation of scenarios and anomaly detection.           \\ \hline
 SDN\&NFV-(E16)  & Passive monitoring & Sniffing packets for analysis. &  Number of malformed packets, header alterations, waypointing violations, and packet re-transmissions. &    Data analytic.          \\
 SDN\&NFV-(E18)  & Offline modeling & Creation of a model based on network topology and SDN controller configuration \cite{Weiwei2024MLbased}.  &  Round-trip time, switch-to-controller traffic, and
controller-to-controller traffic. &    Anomaly detection.          \\
 SDN\&NFV-(E20)  & Offline modelling & Verification of firewall configuration.                       & Number of blackholes and path violations (entire or partial).   &   Network modeling and creation of scenarios.               \\ 
  SDN\&NFV-(E21)  & Active monitoring & Injection of service requests. & Relative CPU usage, memory usage ratio, ratio of service requests, and latency to treat service requests.  & Data analytic. \\
  SDN\&NFV-(E22)  & NDT & Creation and testing of migration scenarios.                       &  Downtime of service, UE recovery time, and latency of service.  &   Creation of scenarios and anomaly detection.            \\ \hline
 AI/ML-(E26)    & NDT & Creation of scenarios and testing of data capturing frameworks.   & Accuracy, end-to-end latency, and total data disclosure incidents.  & Data analytic.              \\ 
 AI/ML-(E28)    & Formal verification & Representation of the AI/ML model placement using a tree graph~\cite{DOro2022OrchestRAN}.   &   Accuracy, end-to-end latency, and number of conflicting decisions.   &   Formulation and solution of the optimization problem.           \\ 
AI/ML-(E30) &  Active monitoring & Data retrieval requests.  & End-to-end latency, accuracy, and attack success rate.  & Creation of attack scenarios.            \\ 
 AI/ML-(E33)   & Offline modelling & Variogram for feature sensitivity analysis.   & Accuracy and end-to-end latency. & Anomaly detection.   \\ \hline
\end{tabular}
\end{table*}

\subsubsection{Active monitoring}
\textcolor{black}{This strategy involves interacting with the system by sending  synthetic service requests or probe packets to uncover any misconfigurations within it. For example, to discover enabled default ports in O-RAN (misconf. I\&O-(E1) in Table \ref{tab4}: enabled default ports, services, accounts, and privileges)}, a series of service requests to the target ports can be produced. The same method can be used to detect deactivated security protocols \textcolor{black}{(misconf. I\&O-(E6) in Table \ref{tab4}: disabled or improper configuration of security protocols to protect reference points)}, such as TLS for A1, by sending and evaluating synthetic connection requests. \textcolor{black}{Note that} most integration and security function misconfigurations (\textcolor{black}{I\&O in} Table \ref{tab3}) can be addressed by active monitoring.  In these cases, data analytics may be utilized to process large amounts of data.

\textcolor{black}{Furthermore, periodically sending probe packets across the network  helps acquire} network status metrics such as latency and bandwidth. This method increases network traffic and only detects potential misconfigurations \textcolor{black}{once they have already affected the system}, indicating a reactive approach~\cite{PARK2023Technology}. \textcolor{black}{For example, consider the misconfiguration I\&O-(E10) in Table \ref{tab4} (sub-optimal equilibrium between FH protection and performance). To detect this misconfiguration, a number of packets can be periodically sent to check any inconsistency in the expected end-to-end latency and throughput (based on SLAs). In this setup, the use of AI/ML models enables the detection of anomalies in measurements.} 

\subsubsection{Passive monitoring}
In contrast to active monitoring, this method involves the study of system elements without the use of probe packets. These solutions often employ sniffer tools for real-time telemetry. For example, \textcolor{black}{the message flow within the  \ac{Near-RT-RIC} interfaces} (A1, E2, O1, Y1) can be monitored to identify protocol misconfigurations \textcolor{black}{(misconf. I\&O-(E2) in Table \ref{tab4}: lack of conformance or interoperability with standard procedures), such as xApp registration/deregistration with the Near-RT RIC}. AI/ML-based analytics may be used to detect abnormalities in real-time telemetry and identify these misconfigurations \cite{PARK2023Technology}. It should be noted that this approach is reactive.

\subsubsection{Formal verification} 
In this approach, the system is formalised using symbolic methods, such as geometry and set theory, and verification techniques are used to detect misconfigurations. These methodologies can provide rigorous evidence of configuration conformance or violation. However, due to the huge scale of the O-RAN system, these verification approaches may be too expensive. Furthermore, verification delays might result in a substantial time gap during which the network may face lower performance and greater exposure to security attacks \cite{Pan2023Misconfiguration}. The combination of AI/ML and formal approaches allows for the speedy and verified discovery of misconfigurations, as demonstrated in~\cite{Oqaily2022MLFM}. \textcolor{black}{For example, to find abnormalities in the generation of policies in O-RAN (misconf. I\&O-(E12) in Table \ref{tab4}: sub-optimal rule generation) (e.g., A1 policies), the rule generation can be represented using a minimal interval set model \cite{Pan2023Misconfiguration}. AI/ML can be used to learn the correlation between this representation and the associated problems (e.g., redundant rules).}

\textcolor{black}{The same approach can be applied to detect  misconfiguration AI/ML-(E28) from Table \ref{tab4} (misplacement of AI/ML model object). However, in this case, a tree graph can serve to model the deployment of a set of AI/ML models in an O-RAN system. Then, a formulation based on binary integer linear programming (BILP) can be applied to find the correlation between the proper placement of the AI/ML models and the system's performance (e.g., accuracy of the AI/ML models, end-to-end latency, etc.)~\cite{DOro2022OrchestRAN}.}

\subsubsection{Offline modeling}
This approach entails parsing the network configuration to provide a quantitative model of the network, enabling the proactive detection of misconfigurations that might compromise meeting \ac{SLA} goals. Using configuration files (logs and configuration databases) as training sets, AI/ML models may learn basic specs. This approach allows for the discovery of SLA violations in the offline model before they occur in the actual implementation. It should be noted that this technique falls short of recording dynamic traffic fluctuations, resulting in the overlooking of some SLA violations~\cite{Oqaily2021SegGuard}. \textcolor{black}{ Consider, for example,  misconfiguration SDN\&NFV-(E18) in Table \ref{tab4} (sub-optimal controller placement). In this example,} the network topology configuration and specifics of SDN controllers (number of controllers, location, and design) can serve as inputs to an ANN to predict system performance (e.g., throughput and latency) \cite{Weiwei2024MLbased}. Based on this model, the SDN placement that causes performance degradation can be identified.

\subsubsection{Network Digital Twin (NDT)}
In this approach, a live virtual representation of O-RAN enables a variety of actions, including emulations, testing, optimisation, monitoring, and analysis of novel configurations in a risk-free environment. This reduces the need for real network deployment, resulting in a proactive approach~\cite{Li2022RLOps}. Using an NDT of the RIC, for example, enables testing of multiple xApps to assess performance and discover any conflicts \textcolor{black}{(see misconfiguration I\&O-(E15) in Table \ref{tab4}: conflicting access to radio resources by xApps). In this example,  AI/ML can help generate conflicting scenarios. The same approach can be applied to detect misconfiguration SDN\&NFV-(E22) in Table \ref{tab4} (sup-optimal service migration), where NDT can be used to create migration scenarios and to detect potential anomalies on monitored KPIs (e.g., inadequate downtime and latency of services).}  It should be noted that the implementation of \textcolor{black}{NDT} necessitates massive resources in terms of storage, computation, maintenance, and the precision required by the models.  AI/ML can help improve the efficiency and precision the of simulate networks and scenarios

\textcolor{black}{It is worth noting that while identifying misconfiguration issues,  not every issue necessitates the use of AI/ML approaches}. Some integration and operation (I\&O) misconfiguration concerns in Table \ref{tab4} can, for example, be automated without the use of AI/ML, such as calculating the number of open ports and system default accounts and passwords. Nonetheless, due to the vast number of components and interfaces in an O-RAN system, data analytics might be useful in discovering configuration issues in massive datasets.

\textcolor{black}{Furthermore, it is important to recognize} that some KPIs in Table \ref{tab4}, such as end-to-end latency and throughput, might signify distinct misconfiguration issues. As a result, tracing back to the source of the misconfiguration to establish the precise misconfiguration type to ease remediation is a major difficulty.

\textcolor{black}{Finally, given the diversity of misconfigurations, different detection approaches may be better suited for certain misconfiguration instances. Therefore, integrating detection approaches into a unified tool can facilitate misconfiguration detection, classification, root cause analysis, and reporting. This system would most likely employ AI/ML to automate and orchestrate the diagnostic process. }

\subsection{Case study: Detection of conflicting xApps}\label{sec:use_case}
\textcolor{black}{This section analyses conflicting xApps. Initially, this misconfiguration problem and its impact in the RAN is described. Then, a detection framework based on insights from prior research is provided. }

\subsubsection{Problem description}
\textcolor{black} {Differing from previous RAN generations,} in a multivendor O-RAN environment, \textcolor{black}{\ac{Near-RT-RIC} xApps}  maintain a high level of independence in \textcolor{black}{their} optimisation or learning process, with only essential data shared \textcolor{black}{between them. In this sense, xApp developers assume direct and isolated management of the \ac{RAN}. This condition may result in numerous overlooked conflicts arising during the combined operation of the xApps within the O-RAN system.} 

\begin{figure}
\centering
\includegraphics[width=\linewidth]{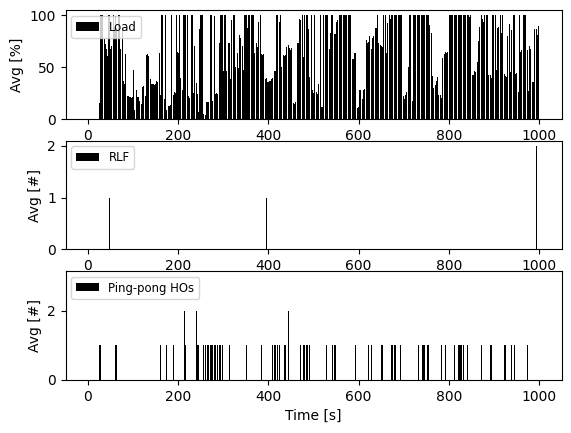}
    \caption{Conflicting MLB and MRO.  The dataset presented in \cite{Adamczyk2023Conflict} has been used to illustrate the issue of ping-pong handovers (HOs) between two apps with conflicting objectives. For clarity, the KPIs of 2 gNBs are shown (the original dataset contains data for 19 gNBs). The MLB maintains the balance of the load on the gNBs (top plot), while the MRO maintains the RLFs close to zero (middle plot). The interaction of these xApps causes multiple ping-pong handovers as illustrated in the bottom plot.} \label{fig:mlb-mro-operation}
\end{figure}

\textcolor{black}{Figure \ref{fig:mlb-mro-operation} exemplifies xApps} causing conflicts using the well-known \ac{SON} functions, notably \ac{MLB} and \ac{MRO}. \ac{MLB} balances traffic distribution among cells to optimise network performance, while MRO ensures robust and stable links to UEs.  Both apps change handover settings, resulting in ping-pong handovers~\cite{Liu2010ConflictAvoidance}.  

Conflicting xApps, like other misconfiguration issues discussed in this study, can have a direct or indirect impact on the O-RAN. For direct impact, conflicts between \ac{RRM} choices made by different xApps result in poor performance and network instability. In terms of indirect impact, the lack of a conflict resolution system for xApps exposes the O-RAN system to security threats.  That is, rogue xApps might use this condition to launch a DoS attack using competing RRM options.

\textcolor{black}{It is important to highlight that solutions designed to address conflicts between applications in the context of \ac{SON} in 4G may not be directly applicable to O-RAN. Specifically, in the approaches involving collaboratively optimising and distributing resources \cite{Huang2018AConflict} or team learning \cite{Zhang2022Team}, the underlying assumption is that all applications are developed by a single vendor with a comprehensive understanding of the interactions among applications and \ac{RAN} elements. However, as previously stated, this scenario may not apply to \ac{O-RAN}. Therefore, a detection of xApp conflicts customized  for O-RAN is required.} 

\subsubsection{Detection approach}

\textcolor{black}{The first step in detecting this misconfiguration issue is to comprehend its nature and construct a model. In this regard, the O-RAN WG3 identified three types of conflicts that may arise in O-RAN:} direct, indirect, and implicit~\cite{ORANWG32023RICARCH}. Table \ref{tab5} illustrates a basic model for these conflicts. In direct conflicts, \textcolor{black}{two (or more)  xApps,  xApp A and xApp B, attempt to operate on the same set of parameters $P_A$ and $P_B$ (i.e., $P_A=P_B$), impacting the same system functions ($I_S$).  Different parameters are changed in indirect conflicts (i.e., $P_A \neq P_B$), yet the impact on the same system functions is represented in the system. Finally, in implicit conflicts, different parameters (i.e., $P_A \neq P_B$) are operated on and different system functions are influenced.} Particularly, implicit conflicts are challenging to solve since the xApps causing the system impact are not known a priori.

\begin{table*}[]
\centering
\caption{Model of conflicts between xApps. \textcolor{black}{$\textit{op}(P_A)$: operation (change, modification) on the set of parameters $P_A$.} }\label{tab5}
\begin{tabular}{p{1.8cm}  p{1.3cm} p{1.3cm} p{3.2cm} p{3.2cm} p{4cm}}
\hline
\rowcolor[HTML]{C0C0C0} 
  &  xApp A & xApp B & Direct conflict & Indirect conflict & Implicit conflict \\ \hline           Set parameters & $P_A$   & $P_B$ & $P_A=P_B$ & $P_A \neq P_B$  &  $P_A \neq P_B$  \\ \hline
  System impact & $P_A  \rightarrow i_A$ &  $P_B  \rightarrow I_B$ & \begin{tabular}[c]{@{}l@{}} \textcolor{black}{ $\textit{op} (P_A) \cap \textit{op}(P_B) \rightarrow I_S$} \\ $I_S = I_A = P_B$  \end{tabular}  & \begin{tabular}[c]{@{}l@{}} \textcolor{black}{$\textit{op}(P_A) \cap \textit{op}(P_B) \rightarrow I_S$} \\ $I_S = I_A = P_B$  \end{tabular} & \begin{tabular}[c]{@{}l@{}} \textcolor{black}{$\textit{op}(P_A) \cap \textit{op}(P_B) \rightarrow I_S$} \\  $I_S \neq I_A  \text{ and } I_S \neq P_B$ \end{tabular} \\ \hline
  Observation &  & & It's known (a priori) which xApps caused $I_S$. & It's known (a priori) which xApps caused $I_S$. & The xApps causing $I_S$ are unknown (a priori). \\ \hline
  Detection &  & & Identify xApps involved in the recent actions (logs). & Identify xApps involved in the recent actions (logs). & Use \ac{AI/ML}, e.g., MDP, to identify the xApps involved. \\ \hline
  \begin{tabular}[c]{@{}l@{}} Example \\ \\ \end{tabular}  & & & \begin{tabular}[c]{@{}l@{}}  Firewall rules: \\ $P_A = \{\text{Allow/Deny UE1}\}$ \\ $P_B = \{\text{Allow/Deny UE1}\}$ \\ $I_S = \text{ Granting UE1}$ \end{tabular} & \begin{tabular}[c]{@{}l@{}} MRO and MLB: \\ $P_A = \{H, TTT\}$ \\ $P_B = \{CIO\}$ \\ $I_S=\text{ Handover boundary}$ \end{tabular} & \begin{tabular}[c]{@{}l@{}} Non-explainable AI/ML-based \\xApps:  \\ $I_S = \text{Delayed impact}$ \end{tabular} \\ \hline
\end{tabular}
\end{table*}

\textcolor{black}{The next step is to design a system that detects the conflicts presented in Table \ref{tab5}. In this context, while recent efforts have contributed to potential conflict detection and mitigation frameworks inside the O-RAN system \cite{Adamczyk2023Conflict}, these efforts have been focused on certain conflict types, and further improvements are necessary to prevent suboptimal outcomes. In addition, a critical concern in the design of conflict detection is to provide a generalised detection solution for the three categories of conflicts, if possible. }

\begin{figure*}
\centering
\includegraphics[width=.85\linewidth]{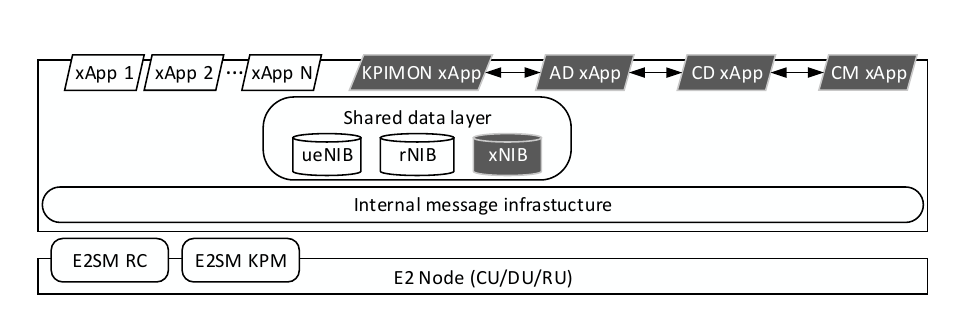}
    \caption{Managing conflicts between xApps: detection and mitigation using \ac{AI/ML} techniques. Shaded components have been incorporated into the original \ac{Near-RT-RIC} architecture of the \ac{O-RAN} WG3 \cite{ORANWG32023RICARCH}, which include the information database for xApp actions (xNIB) and the xApps KPIMON, AD, CD, and CM.} \label{fig:Conflicts_Identification_framework}
\end{figure*}

Figure \ref{fig:Conflicts_Identification_framework} presents \textcolor{black}{our} proposed framework that helps identify conflicts across xApps, based on the standard \ac{Near-RT-RIC} architecture \cite{ORANWG32023RICARCH} and earlier research \cite{Adamczyk2023Conflict}. \textcolor{black}{Note that the automated mitigation of the detected conflicts is also considered}.  The conflict detection and mitigation functionalities in this framework are implemented as xApps, notably CD xApp and CM xApp. Furthermore, this framework uses other xApps, particularly KPIMON xApp and AD xApp~\cite{ORANSC2023RICAPP}, and establishes a new network information database, referred to as xNIB, to store the \textcolor{black}{operations of the xApps}.

\begin{figure}
\centering
\includegraphics[width=\linewidth]{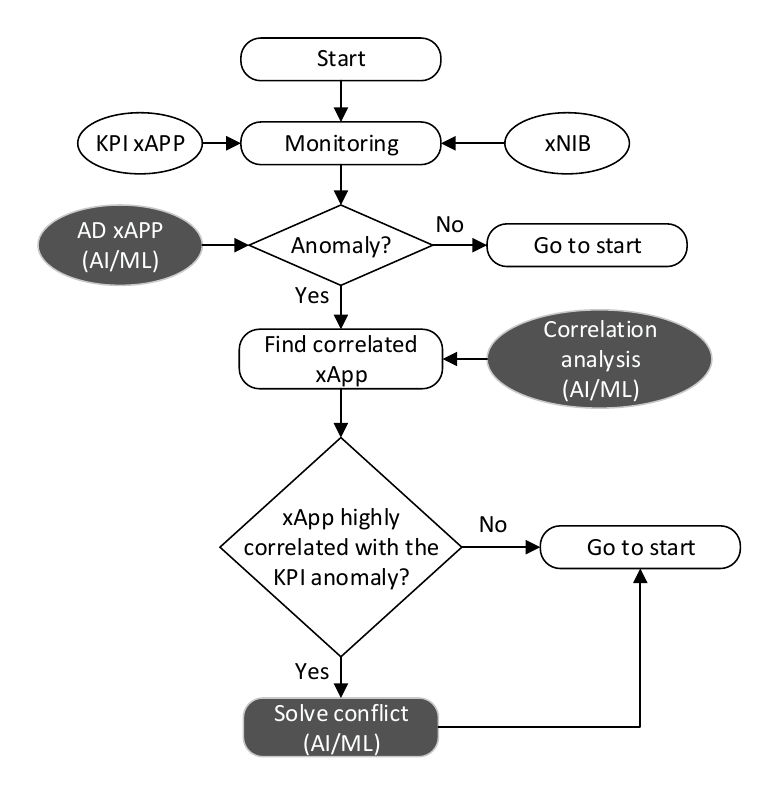}
    \caption{Method for detection and mitigation of conflict xApps. Shaded components may require \ac{AI/ML} techniques.} \label{fig:Conflicts_Identification_Algorithm}
\end{figure}

In the method outlined in Figure \ref{fig:Conflicts_Identification_Algorithm}, both the KPIs of the E2 nodes and the activities of the xApps are monitored. It should be noted that the KPIs are dependent \textcolor{black}{on the use case of the xApps}. \textcolor{black}{For example,} the mean load of the base station, the number of call blockages, the number of radio connection failures, and the number of handovers,  may be monitored to discover conflicts between MLB and MRO xApps \cite{Adamczyk2023Conflict}. \textcolor{black}{The anomaly detection (AD) xApp}  evaluates the variability of the KPIs collected by the KPIMON xApp. If a considerable drop in system performance is noticed, the system investigates the actions of xApps (consults the xNIB) to determine the source of the dispute.

In cases of implicit conflicts, more advanced correlation processes, such as \ac{MDP} and Bayesian models, may be required to identify which xApps are generating the conflicts. Once the conflicting xApps have been found, they may be blocked directly based on priority. However, as seen in \cite{Adamczyk2023Conflict}, these strategies may provide suboptimal outcomes. As a result, \ac{RL} can be used to learn the best way to assign priority in order to resolve conflicts and maximise system efficiency.  Note that due to the complexities of O-RAN management, at least three components in Figure \ref{fig:Conflicts_Identification_Algorithm} use AI/ML.

 \textcolor{black}{Note that at the time of writing this document, none of the existing open-source or commercial \ac{Near-RT-RIC}s have incorporated a conflict detection and mitigation solution. We consider that experimenting with this approach of conflict detection among xApps in an O-RAN testbed could represent a significant milestone in the field of misconfiguration solutions. This will promote the development of multi-vendor xApps for optimizing RAN operations while also guaranteeing reliability and minimizing outages. However, conducting such testing is beyond the scope of this analytical study.}  

\subsection{Summary and insights}
Although we examined how AI/ML can be used to detect misconfigurations using various detection methodologies, we emphasize that not all misconfiguration issues necessitate the use of AI/ML-based detection approaches. Nonetheless, the introduction of AI/ML can help simplify the analysis of the metrics captured in O-RAN deployments with a large number of components, applications, and amounts of traffic.

There are a number of commercially available tools (e.g., \cite{Viavi2024TeraVM}, \cite{Keysight2024P8828S}, and \cite{Spirent2024ORAN}) that offer misconfiguration detection. Their documentation indicates the use of both passive and active monitoring to identify integration and security function misconfigurations. Detection tools for the other misconfiguration issues have yet to be developed. Those associated with AI/ML rely on specific application use-cases.

Furthermore, three types of conflicts are considered in the strategy illustrated in the case study. However, although there are several examples of xApps that cause direct and indirect conflicts, to the best of our knowledge, no examples of implicit conflicts have been published. These conflicts are predicted to arise when more AI/ML-powered xApps are added to the O-RAN, particularly if the xApps utilise complex DNNs (non-explainable AI/ML). 
  
The biggest challenge in studying the detection of misconfiguration is that O-RAN technology is still in its early phases of development, making it difficult to evaluate misconfiguration issues in real deployments. Existing experimental testbeds are rather simple. For example, they include just the \ac{Near-RT-RIC} but not the \ac{Non-RT-RIC}. As a result, datasets relating to O-RAN misconfigurations are unavailable. Additional efforts are required to produce these materials, allowing the research community to analyze O-RAN misconfigurations and to suggest and test mitigation methods.

\section{Conclusion} \label{sec:5}

O-RAN characteristics such as disaggregation, openness, and intelligence provide exciting opportunities for innovation in 5G and 6G networks. However, as illustrated in this study, these characteristics may cause misconfiguration issues that can significantly impact on the security and performance of the system.  

As the O-RAN develops, certain methods for detecting misconfigurations associated with system integration and operation are emerging. However, use case-specific misconfiguration problems have yet to be explored. For instance, most distributed AI/ML implementations (model sharing, model splitting, and federated learning) are yet to be validated. In this work, we have highlighted the AI/ML-related misconfiguration issues that must be addressed so that the benefit of intelligence in the \ac{O-RAN} is realised, rather than the intelligence becoming a limiting factor or a source of exploitation.

\section*{Acknowledgments}
This work is supported through the \textcolor{black}{NICYBER2025 programme funded by Innovate UK. The ORANSecAI project is a collaboration with Ampliphae.} The views expressed are those of the authors and do not necessarily represent the project or the funding agency.



\printcredits

\bibliographystyle{IEEEtran}

\bibliography{cas-refs}



\end{document}